\documentclass[journal]{IEEEtran}
\usepackage{amsmath}
\usepackage{amsfonts}
\usepackage{algorithmic}
\usepackage{color}
\usepackage{array}
\usepackage{algorithmic}
\usepackage[linesnumbered,ruled,vlined]{algorithm2e}
\usepackage[caption=false,font=normalsize,labelfont=sf,textfont=sf]{subfig}
\usepackage{times}
\usepackage{graphicx}
\usepackage{longtable}
\begin{document}
\makeatletter
\newcommand{\nosemic}{\renewcommand{\@endalgocfline}{\relax}}
\newcommand{\dosemic}{\renewcommand{\@endalgocfline}{\algocf@endline}}
\newcommand{\pushline}{\Indp}
\newcommand{\popline}{\Indm\dosemic}
\let\oldnl\nl
\newcommand{\nonl}{\renewcommand{\nl}{\let\nl\oldnl}}
\makeatother
\title{Study of Enhanced Subset Greedy Multiuser Scheduling for Cell-Free Massive MIMO Systems} \vspace{-0.75em}

\author{Saeed Mashdour$^{\star}$, Rodrigo C. de Lamare $^{\star,\dagger}$ and Joao P. S. H. Lima $^{\ddagger}$
\thanks{The authors $^{*}$ are with the Centre for Telecommunications Studies,
Pontifical Catholic University of Rio de Janeiro, Brazil,
$^{\dagger}$ is with the University of York, UK, and $^{\ddagger}$
is with CPqD, Campinas, Brazil. Emails: smashdour@gmail.com,
delamare@cetuc.puc-rio.br, jsales@cpqd.com.br. This paper was
supported by CNPQ, FAPERJ, FAPESP - Grant No. 2015/24499-0, CPqD and
Funttel/Finep - Grant No. 01.20.0179.00.}}

\maketitle

\begin{abstract}
In this work, we consider the problem of multiuser scheduling for
the downlink of cell-free massive multi-input multi-output networks
with clustering. In particular, we develop a multiuser scheduling
algorithm based on an enhanced greedy method that is deployed with
linear precoding and clustering. Closed-form expressions for the
sum-rate performance are derived when imperfect channel state
information is considered. The proposed scheduling algorithm is then
analyzed along with its computational cost and network signaling
load. Numerical results show that the proposed scheduling method
outperforms the existing methods and in low signal-to-noise ratios,
its performance becomes much closer to the optimal approach.
\vspace{-0.5em}
\end{abstract}

\vspace{-0.75em}

\section{Introduction}
Cell-free (CF) massive multi-input multi-output (MIMO) networks were
introduced in \cite{CellFree} as an architecture with a large number
of single-antenna access points (APs) serving a smaller number of
users in the same time-frequency resource that can offer higher
throughput and coverage as compared to multi-cell networks. In such
networks, all the APs are connected to a central processing unit
(CPU) which is responsible for coordination and processing the
signals of the users and the precoding which is considered for
performance improvement
\cite{mmimo,wence,Robust,gbd,mbthp,wlbd,cqabd,siprec,bbprec,okspme,lrcc,rsrbd,rsthp,rapa,cesg,jidf,dynovs,spa,tds,mbdf,baplnc,memd}
However, a CF network requires a huge signaling load and computational cost which is not practical. Therefore, clustering including network-centric and user-centric \cite{Scalable Cell-Free} and \cite{Clustered Cell-Free} approaches have been adopted to reduce the cost so that each user is only served by a subset of APs.

Since all the APs in CF massive MIMO networks use the same time-frequency resource for communication, multiuser interference is unavoidable. 
Multiuser scheduling could reduce multiuser interference by selecting a set of users with lowest spatial correlation \cite{Ubiquitous} in each transmission, improving the network performance. In \cite{Distributed user}, a distributed multiuser scheduling algorithm according to the virtual signal to interference and noise ratio (SINR) of the users is proposed in multiuser MIMO networks and implies multiuser scheduling as an important factor to achieve a desirable sum-rate and spectral efficiency. Multiuser scheduling has recently attracted a lot of attention in CF massive MIMO systems. The work of \cite{Improving Cell-Free} introduced a framework in CF massive MIMO networks to partition users into groups according to their mutual interference level and proposed a semidefinite relaxation algorithm to schedule the groups on different radio resources, mitigating the effect of inter-group interference. The work in \cite{Resource Allocation} has solved a weighted sum-rate (WSR) problem using fractional programming and employed compressive sensing for multiuser scheduling in a clustered CF massive MIMO network {while imperfect CSI is considered and the perfect CSI is treated as a special case}.

{In this paper, we investigate the downlink of CF and clustered CF massive MIMO networks and propose a clustered enhanced subset greedy (C-ESG) multiuser scheduling algorithm which takes a significant step beyond that of existing greedy techniques by introducing a multiple user set strategy along with an evaluation of the sets, which results in a performance close to the optimal exhaustive search. We analyze C-ESG and assess its computational cost in both CF and clustered CF networks. Closed-form expressions for the sum-rate are derived and the performance of C-ESG is compared with existing scheduling techniques. Simulations show that C-ESG outperforms existing techniques and the effect of clustering on the performance, complexity and signaling load.}

{\it Notation}: In the paper, $\textbf{I}_{n}$ denotes the $n\times n$ identity matrix, the complex normal distribution is represented by $\mathcal{CN}\left ( .,. \right )$, superscripts $^{T}$, $^{\ast}$, and $^{H}$ denote transpose,
complex conjugate and hermitian operations respectively, $\mathcal{A}\cup \mathcal{B} $ is union of sets $\mathcal{A}$ and $\mathcal{B}$, and $\mathcal{A}\setminus \mathcal{B} $ shows exclusion of set $\mathcal{B}$ from set $\mathcal{A}$. \vspace{-0.85em}

\section{System Model}

We consider the {downlink of} a CF massive MIMO network with $M$ single-antenna access points (APs) and $K$ uniformly distributed single-antenna users {so that the total number of users in the network is much larger than the number of APs $K>>M$}. We employ a 
clustering approach which divides the whole area into $C$ 
{non-overlapping equal size areas each including the APs and the users of that area that form a cluster.} \vspace{-0.75em}
\vspace{-0.75em}
\subsection{Cell-Free Network}\label{AA}

The channel coefficient between the $m$th AP and the $k$th user is
denoted by $g_{m,k}=\sqrt{\beta _{m,k}}h_{m,k}$  where $\beta
_{m,k}$ shows the large scale fading coefficient and $ h_{m,k}\sim
\mathcal{CN}\left ( 0,1 \right )$ shows the small-scale fading
coefficient, defined as independent and identically distributed
(i.i.d) random variables (RVs) that remain constant during a
coherence interval and are independent over different coherence
intervals \cite{CellFree}. The received signal in the downlink
transmission is given by \vspace{-0.75em}
\begin{equation} \label{eq:CF-sig}
    \textbf{y}=\sqrt{\rho _{f}}\textbf{G}^T\textbf{P}\textbf{x}+\textbf{w},
\end{equation}
where $\rho _{f}$ is the maximum transmitted power of each antenna,
$\textbf{G}\in \mathbb{C}^{M\times K}$  is the channel matrix with
elements $\left [ \textbf{G} \right ]_{m,k}=g_{m,k}$,
$\textbf{w}=\left [ w_{1},\cdots,w_{K}  \right ]^{T}$ is the
additive noise vector with $\textbf{w}\sim \mathcal{CN}\left (
0,\sigma_{w}^{2}\textbf{I}_{K} \right )$, $\textbf{P}\in
\mathbb{C}^{M\times K}$ is the linear precoder matrix such as those
originated from minimum mean-square error (MMSE) or zero-forcing
(ZF) designs, $\textbf{x}=\left [ x _{1},\cdots ,x_{K} \right ]^{T}$
is the zero mean symbol vector which is independent of noise and
channel coefficients and its elements are mutually independent
$\textbf{x}\sim \mathcal{CN}\left ( \textbf{0},\textbf{I}_{K} \right
)$. In the presence of imperfect channel state information (CSI),
equation (\ref{eq:CF-sig}) is rewritten as \vspace{-2mm}
\begin{multline}
\textbf{y}=\sqrt{\rho _{f}}\left ( \hat{\textbf{G}}+\tilde{\textbf{G}} \right )^T\textbf{P}\textbf{x}+\textbf{w}\\
=\sqrt{\rho _{f}}\hat{\textbf{G}}^{T}\textbf{P}\textbf{x}+\sqrt{\rho _{f}}\tilde{\textbf{G}}^{T}\textbf{P}\textbf{x}+\textbf{w},
\end{multline}
where $\hat{\textbf{G}}\in \mathbb{C}^{M\times K}
$ is the channel matrix estimate and $\tilde{\textbf{G}}\in \mathbb{C}^{M\times K}$ is the estimation error matrix. 
{Assuming Gaussian signaling, the upper bound on the achievable sum-rate of the CF system with imperfect CSI is given by
} \vspace{-0.5em}
\begin{equation}\label{eq:RCF}
    R_{CF}=\log_{2}\left (  \det\left [\textbf{R}+\textbf{I}_K  \right ]\right )
\end{equation}
where the covariance matrix $\textbf{R}=E[\textbf{y}\textbf{y}^H ]$ is expressed by \vspace{-0.75em}
\begin{equation}\label{eq:RCF_1}
    \textbf{R}=\rho _{f} \hat{\textbf{G}}^{T}\textbf{P}\textbf{P}^{H}\hat{\textbf{G}}^{\ast }\left ( \rho _{f}\tilde{\textbf{G}}^{T}\textbf{P}\textbf{P}^{H}\tilde{\textbf{G}}^{\ast } +\sigma _{w}^{2}\textbf{I}_K\right )^{-1}
\end{equation}
and $\textbf{x}$ and $\textbf{w}$ are statistically independent.
\vspace{-4mm}
\subsection{Clustered Cell-Free Network}\label{BB}

\begin{figure}
    \centering
        \includegraphics[width=.65\linewidth]{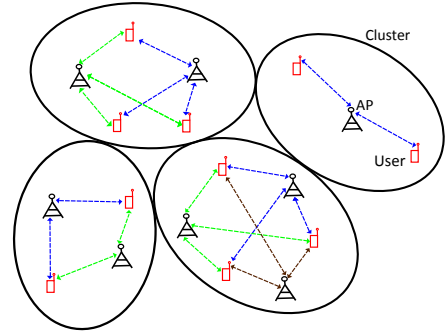}
    \vspace{-0.85em}
    \caption{{Clustered CF network.}}
    \label{fig:Clusters}
\end{figure}

We consider a CF network with $C$ clusters, 
as illustrated in Fig.~\ref{fig:Clusters}. The signal received by the users of the $c$th cluster is given by  \vspace{-0.5em}
\begin{multline} \label{eqQ}
\textbf{y}_{c}=\sqrt{\rho _{f}}\left ( \hat{\textbf{G}}_{cc}+\tilde{\textbf{G}}_{cc} \right )^T\textbf{P}_{c}\textbf{x}_{c}+\\
\sum_{i=1,i\neq c}^{C}\sqrt{\rho _{f}}\left ( \hat{\textbf{G}}_{ic}+\tilde{\textbf{G}}_{ic} \right )^T\textbf{P}_{i}\textbf{x}_{i}+\textbf{w}_{c}=\\
\sqrt{\rho _{f}}\hat{\textbf{G}}_{cc}^T\textbf{P}_{c}\textbf{x}_{c}+\sqrt{\rho _{f}}\tilde{\textbf{G}}_{cc}^T\textbf{P}_{c}\textbf{x}_{c}+\sum_{i=1,i\neq c}^{C}\sqrt{\rho _{f}}\hat{\textbf{G}}_{ic}^T\textbf{P}_{i}\textbf{x}_{i}+\\
\sum_{i=1,i\neq c}^{C}\sqrt{\rho _{f}}\tilde{\textbf{G}}_{ic}^T\textbf{P}_{i}\textbf{x}_{i}+\textbf{w}_{c}
\end{multline}
where $\hat{\textbf{G}}_{ic}$, $\tilde{\textbf{G}}_{ic} \in \mathbb{C}^{M_i\times K_c}$ are respectively channel estimation and estimation error matrices from APs of the cell $i$ to users of the cell $c$, $\textbf{P}_{i}\in \mathbb{C}^{M_i\times K_c}$ is the linear precoding matrix, and  $\textbf{x}_i=\left [ x_{i1},\cdots ,x_{iK_c} \right ]^{T}$, $\textbf{x}_i\sim \mathcal{CN}\left ( \mathbf{0},\mathbf{I}_{K_{c}} \right )$ is the symbol vector of the cluster $i$, $i\in \left \{ 1, 2, \cdots ,C \right \}$. 
{Accordingly, the upper bound on the achievable sum-rate of the clustered network is} \vspace{-0.5em}
\begin{equation}\label{eq:RCL}
    R_{cl}=\sum_{c=1}^{C}R_{c}
\end{equation}
\vspace{-2mm}
{where the upper bound on the sum-rate in cluster c is}
\begin{equation}\label{eq:RCL0}
    R_{c}=\log_{2}\left (  \det\left [\left ( \rho _{f}\hat{\textbf{G}}_{cc}^T\textbf{P}_{c}\textbf{P}_{c}^{H}\hat{\textbf{G}}_{cc}^* \right )\textbf{R}_{c}^{-1}+\textbf{I}_{K_c}  \right ]\right )
\end{equation}
and the covariance matrix $\textbf{R}_{c}$ is described by
\begin{multline}\label{eq:RCL1}
\textbf{R}_{c}=E\left [ \left ( \textbf{y}_{c}-\sqrt{\rho _{f}}\hat{\textbf{G}}_{cc}^T\textbf{P}_{c}\textbf{x}_{c} \right )\left ( \textbf{y}_{c}-\sqrt{\rho _{f}}\hat{\textbf{G}}_{cc}^T\textbf{P}_{c}\textbf{x}_{c} \right )^{H} \right ]\\
=\rho _{f}\tilde{\textbf{G}}_{cc}^T\textbf{P}_{c}\textbf{P}_{c}^{H}\tilde{\textbf{G}}_{cc}^*+\sum_{i=1,i\neq c}^{C}\rho _{f}\hat{\textbf{G}}_{ic}^T\textbf{P}_{i}\textbf{P}_{i}^{H}\hat{\textbf{G}}_{ic}^*+\\
\sum_{i=1,i\neq c}^{C}\rho _{f}\tilde{\textbf{G}}_{ic}^T\textbf{P}_{i}\textbf{P}_{i}^{H}\tilde{\textbf{G}}_{ic}^*+\sigma _{w}^{2}\textbf{I}_{K_c}
\end{multline}
and $\textbf{x}_c$ and $\textbf{w}_c$ are statistically independent. 

\vspace{-2mm}
\section{Proposed Clustered Enhanced Subset Greedy Algorithm}
{Since $M$ single-antenna APs cannot serve $K$ single-antenna users, we have to find a smaller set of users which satisfies a system performance criterion}. We can use an exhaustive search that compares performance of the all possible sets of users and selects the best set in each cluster. However, an exhaustive search has a huge computational cost and is thus impractical. Greedy algorithms are important mathematical techniques which have low cost and are simple and effective at approaching the global optimal solutions of complex problems, which motivates us to develop C-ESG. Therefore, based on the greedy algorithm in \cite{Multiuser}, we propose C-ESG which has much lower computational cost while approaching the performance of the optimal exhaustive search. {C-ESG introduces a refined search with multiple sets of users and an evaluation step of the sets that departs from existing greedy techniques and allows a specific number of users to be scheduled.} 

In the proposed C-ESG algorithm, we first find a primary set of users in each cluster by adapting a similar approach to that of \cite{OnDownlink}. We apply the MMSE precoder considering the channel matrix of the users as $\textbf{G}_{cc} \in \mathbb{C}^{M_c\times K_c}$, {where $K_c$ and $M_c$ are the number of users and APs in the intended cluster, respectively}. When the number of receive antennas is larger than the number of transmit antennas $K_c > M_c$, we aim to schedule $n$ users out of $K_c$ users so that $n \leq  M_c$ and we can achieve a desirable sum-rate in those users. The selected set of users is shown by $S_{n}$ and results in a row-reduced channel matrix $\textbf{G}_{cc}\left ( S_{n} \right )$. The goal is to obtain the highest achievable sum-rate as a solution to the problem
\begin{equation}
\begin{aligned}
& \underset{S_{n}}{\text{max}}~R_{MMSE}\left ( S_{n} \right ) \\
& \text{subject to} \ \left \| \textbf{P}_c\left ( S_{n} \right ) \right \|^{2}\leq P.
\end{aligned}
\end{equation}
where $R_{MMSE}\left ( S_{n} \right )$ is defined as the sum-rate with the MMSE precoder when $S_{n}$ is the set of intended users, $P$ is the upper limit of the signal covariance matrix $\textup{Trace}\left [ \textbf{C}_{\textbf{x}}\right ]\leq P$, $\textbf{P}_c\left ( S_{n} \right )=\textbf{W}_{n}\textbf{D}_{n}$ is the precoding matrix including the {normalized MMSE weight matrix} $\textbf{W}_{n}\in\mathbb{C}^{M_c \times n} $ and the power {allocation matrix} $\textbf{D}_{n}$ defined as 
\begin{equation}
    \textbf{D}_{n}=\begin{bmatrix}
\sqrt{p_{1}} & 0 & \cdots  & 0\\
 0& \sqrt{p_{2}} & \cdots &0 \\
 \vdots & \vdots  &\cdots   & \vdots \\
 0&0  &\cdots   & \sqrt{p_{n} }
\end{bmatrix}
\end{equation}
With equal power loading for simplicity and to focus on scheduling, a suboptimal greedy algorithm to solve the problem is used in the 1st stage of Algorithm \ref{alg:alg1}, and the primary user set $S_{n\left ( 1 \right )}$ in each cluster is obtained. Then, {to assess more sets, and in order to identify the minimum sum-rate, we select the user with the lowest channel power from the first selected set $S_{n\left ( 1 \right )}$,} which we call the first excluded user as \vspace{-2mm}
\begin{equation} \label{kex}
         k_{ex\left (1\right )}=\underset{k\in S_{n\left ( 1 \right )}}{{\arg \min}}~ \textbf{g}_{k}^{H}\textbf{g}_{k}
\end{equation}
where $\textbf{g}_{k}\in \mathbb{C}^{M_c \times 1}$ is channel vector to $k$th user. Considering $\mathcal{K}_{cl}=\left \{ 1,2,\cdots ,K_c \right \}$ as set of all users of the cluster, we define the first set of remaining or unselected users of the cluster as $\mathcal{K}_{clr\left ( 1\right )}=\mathcal{K}_{cl}\setminus S_{n\left ( 1 \right )}$, which include the users other than $S_{n\left ( 1 \right )}$.  From the first set of remaining users, we select the user with the highest channel power as the first new user as \vspace{-0.5em}
\begin{equation} \label{knew}
 k_{new\left (1\right )}=\underset{k\in \mathcal{K}_{clr\left ( 1\right )}}{{\arg \max}}~\textbf{g}_{k}^{H}\textbf{g}_{k}
\end{equation}
Substituting the first excluded user by the first new user in $S_{n\left ( 1 \right )}$, we achieve the second set of users as $S_{n\left ( 2 \right )}$. We continue this procedure for the second set and so on, until we get 
{${K_{c}-n}$} user sets in addition to the first set. 
Then, the $i$th user set and the $i$th remaining user set are shown as follows: 
\vspace{-1mm}
\begin{equation}
    S_{n\left ( i \right )}=\left (S_{n\left ( i-1 \right )}\setminus k_{ex\left (i-1\right )} \right )\cup k_{new\left (i-1\right )}
\end{equation}
\vspace{-6mm}
\begin{equation}
    \mathcal{K}_{clr\left ( i\right )}=\mathcal{K}_{clr\left ( i-1\right )}\setminus k_{new\left (i-1\right )}
\end{equation}
 where 
 {$i \in \left \{ 2,\cdots ,{K_{c}-n}+1 \right \}$. 
 In order to select the best set among the acquired sets, we can use the sum-rate expression in (\ref{eq:RCL}) for clustered CF network. Thus, the best set is chosen: 
\begin{equation}
         S_{n_{b}}=\underset{S_{n} \in S_{n\left ( j \right )}}{\textup{argmax}}\left \{  R_{c}\left ( S_{n} \right ) \right \}
\end{equation}
where
 {$j \in \left \{ 1,\cdots ,{K_{c}-n}+1 \right \}$. The details of C-ESG are shown in Algorithm \ref{alg:alg1}. We note that for applying the C-ESG algorithm to the CF system, we would change the $K_{c}$ to $K$ as the number of users in the CF network, 
$\mathcal{K}_{cl}$ to $\mathcal{K}=\left \{ 1,2,\cdots ,K \right
\}$, $\mathcal{K}_{clr\left ( i\right )}$ to $\mathcal{K}_{r\left (
i\right )}$ as the $i$th remaining user set, and accordingly, other
functions and parameters would change to the network-wide level such
as $R_{CF}$ instead of $R_{c}$.
\begin{algorithm}
\LinesNumbered
\SetKwBlock{Begin}{}{}
\caption{Proposed C-ESG Scheduling Algorithm.}\label{alg:alg1}
\SetAlgoLined
stage=1
  \Begin{
  \textbf{set} $l = 1$\;
  \textbf{find a user such that}

  $u_{1}=\underset{k\in \mathcal{K}_{cl}}{\textup{argmax}} ~\textbf{g}_{k}^{H}\textbf{g}_{k}$\;
  \textbf{set $U_1 = {u_1}$ and denote the achieved rate}

  $R_{MMSE}\left ( U_{1} \right )$;

  \While  {$l<n$} {
   \hspace{0.5cm} $l=l+1$;

  \hspace{0.5cm} \textbf{find a user $u_{l}$ such that}

  \hspace{0.5cm} $u_{l}=\underset{k\in\left (  \mathcal{K}_{cl} \setminus  U_{l-1} \right )}{\textup{argmax}}R_{MMSE}\left ( U_{l-1}\cup \left \{ k \right \} \right )$;

  \hspace{0.5cm} \textbf{set $U_{l}=U_{l-1}\cup \left \{ u_{l} \right \}$ and denote the rate}

  \hspace{0.5cm} $R_{MMSE}\left ( U_{l} \right )$;

  \hspace{0.5cm} \textbf{If $R_{MMSE}\left ( U_{l} \right )\leq R_{MMSE}\left ( U_{l-1} \right )$, break}

  \hspace{0.5cm} $l=l-1$;}

  }
\end{algorithm}

\begin{algorithm}
\LinesNumbered
\setcounter{AlgoLine}{15}
\SetKwBlock{Begin}{}{}
\SetAlgoLined
\nonl
  \Begin{
  $S_{n\left ( \textup{stage} \right )}=U_{l}$\;
  \textbf{compute}: $ R_{c}\left ( S_{n\left ( \textup{stage} \right )}\right )$\;
  $\mathcal{K}_{clr\left ( \textup{stage}\right )}=\mathcal{K}_{cl}\setminus S_{n\left ( \textup{stage} \right )}$\;
  $ k_{ex\left (\textup{stage}\right )}=\underset{k\in S_{n\left ( \textup{stage} \right )}}{\textup{argmin}}\textbf{g}_{k}^{H}\textbf{g}_{k}$\;
  $k_{new\left (\textup{stage}\right )}=\underset{k\in \mathcal{K}_{clr\left ( \textup{stage}\right )}}{\textup{argmax}}\textbf{g}_{k}^{H}\textbf{g}_{k}$\;

  }
  \For {$\textup{stage}=2$ \textup{to}
{${K_{c}-n}+1$}} {
$S_{n\left ( \textup{stage} \right )}=\left (S_{n\left ( \textup{stage}-1 \right )}\setminus k_{ex\left (\textup{stage}-1\right )}  \right )\cup k_{new\left (\textup{stage}-1\right )}$\;
$\mathcal{K}_{clr\left ( \textup{stage}\right )}=\mathcal{K}_{clr\left ( \textup{stage}-1\right )}\setminus k_{new\left (\textup{stage}-1\right )}$\;
$ k_{ex\left (\textup{stage}\right )}=\underset{k\in S_{n\left ( \textup{stage} \right )}}{\textup{argmin}}\textbf{g}_{k}^{H}\textbf{g}_{k}$\;
$k_{new\left (\textup{stage}\right )}=\underset{k\in \mathcal{K}_{clr\left ( \textup{stage}\right )}}{\textup{argmax}}\textbf{g}_{k}^{H}\textbf{g}_{k}$\;
\textbf{compute}: $R_{c}\left ( S_{n\left ( \textup{stage} \right )} \right )$;\
  }
$S_{n_{b}}=\underset{S_{n} \in S_{n\left ( j \right )}}{\textup{argmax}}\left \{  R_{c}\left ( S_{n} \right ) \right \}$\;
\textbf{Linear MMSE precoding of n scheduled users}
\end{algorithm}
\section{Analysis of the Proposed C-ESG Algorithm}
In the proposed C-ESG algorithm, we use the channel power of the users so that there are different sets of users and we can {assess more possible sets, approaching the optimal set while the complexity is significantly less than that of the exhaustive search. In each step of the C-ESG algorithm, we drop the user with the lowest channel power and add the user with the highest channel power so that a new set is achieved. If we schedule the maximum possible number of users $n=M_c$, we can show the sets of the {selected} users in stages $\left \{ 1, 2, \cdots, j, \cdots , 
{{K_{c}-M_{c}}+1}  \right \}$ as
\vspace{-3mm}
\begin{multline}
    \mathcal{S}_{EG}=
    \left \{ S_{EG_{1}}, S_{EG_{2}}, \cdots , S_{EG_{j}}, \cdots ,
    {S_{EG_{K_{c}-M_{c}+1}}}\right \}
\end{multline}
where $S_{EG_{j}}$ is the $j$th set of the proposed C-ESG method. Then, for C-ESG, there would be 
{${K_{c}-M_{c}}+1$} sets in each cluster of the clustered CF system and 
{${K-M}+1$} sets in the CF network. For the exhaustive search, we have the following sets as all the possible user sets for all possible stages $\left \{ 1, 2, \cdots, j, \cdots , 
{\frac{K_c!}{M_{c}!\left ( K_c-M_{c} \right )!}}  \right \}$
\vspace{-4mm}
\begin{multline}
   \hspace{-2mm} \mathcal{S}_{Ex}=
    \left \{ S_{Ex_{1}}, S_{E2_{2}}, \cdots , S_{Ex_{i}}, \cdots , 
    {S_{Ex_{\frac{K_c!}{M_{c}!\left (K_c-M_{c} \right )!}}}} \right \}
\end{multline}
where $S_{Ex_{i}}$ is the $i$th set of the exhaustive search method. Therefore, there are 
{$\frac{K_c!}{M_{c}!\left ( K_c-M_{c} \right )!}$}sets in each cluster of the clustered CF network, and 
{$\frac{K!}{M!\left ( K-M \right )!}$} sets in the CF network. Thus, the cost of the C-ESG algorithm is much lower than that of the exhaustive search especially for large $K_c$ or $K$.





















\emph{Proposition}. The sum-rate of C-ESG is bounded as
\vspace{-1.85mm}
\begin{equation}
         R_c\left ( S_{G} \right )\leq R_c\left ( S_{C} \right ) \leq R_c\left ( S_{X} \right )
\end{equation}
where $S_{G}$ is the set selected by the standard greedy method \cite{OnDownlink} as shown by $S_{n}$ in the first stage of the Algorithm \ref{alg:alg1}, $S_{C}$ is the set selected by C-ESG algorithm, and $S_{X}$ is the selected set by the exhaustive search

$S_{C}=\underset{S_{EG_{j}}, j=1:
{{K_c-M_{c}}+1}}{\textup{argmax}}\left \{ R_c \left ( S_{EG_{j}} \right )\right \}$

$S_{X}=\underset{S_{Ex_{i}}, i=1:
{\frac{K_c!}{M_{c}!\left ( K_c-M_{c} \right )!}}}{\textup{argmax}}\left \{ R_c \left ( S_{Ex_{i}} \right )\right \}$

\hspace{-0.36cm}\textbf{Proof.}
Considering
$S_{G}=\left \{ n_{g_{1}},n_{g_{2}},...,n_{g_{M_{c}-1}},
{n_{g_{M_{c}}}} \right \}$ as the first set in C-ESG, suppose that
we are at the second stage and the corresponding set is considered
as $S_{En_{2}}$ 
which is different from $S_{G}$ in one element, if $m_1$ and 
{$n_{g_{M_{c}}}$} provide equal sum-rates, then, we would have
$R_c\left ( S_{G} \right )=R_c\left ( S_{C} \right )$.
In the case that we are in the $j$th stage (including $j=2$), if we can have a better choice of subsets and if there is a subset in $S_{En_{j}}$ that differs from $S_{G}$, then this would result in $R_c\left ( S_{C} \right )=R_c\left ( S_{G} \right )+\epsilon$, where $\epsilon>0$. Thus, we can conclude that  $R_c\left ( S_{C} \right ) \geq R_c\left ( S_{G} \right )$.
On the other hand, according to the combinations for the exhaustive search which includes all the possible cases, $S_{C}$ is a special set of the exhaustive search combinations. Therefore, considering $S_{X}=\left \{ n_{1},n_{2},...,
{n_{M_{c}}} \right \}$, if C-ESG results in the same set $S_{C}=\left \{ n_{1},n_{2},...,
{n_{M_{c}}} \right \}$, then, $R_c\left ( S_{C} \right )= R_c\left ( S_{X} \right )$. However, if the set selected by the exhaustive search is a different set from the selected set by C-ESG, $S_{X}$ would clearly be with a higher sum-rate than $S_{C}$. Thus, we would obtain $R_c\left ( S_{C} \right ) \leq   R_c\left ( S_{X} \right )$. Note that the same proposition holds for the analysis of the ESG algorithm in the CF network.
\vspace{-2mm}
\section{Impact of Clustering on Signaling Load and Scheduling Cost}
\vspace{-1mm}
The network-centric clustering technique divides the APs into non-overlapping cooperation clusters where the APs of each cluster collaborate in serving the users located in their joint coverage area \cite{Networked, Increasing}. Note that extensions to overlapping clusters are also possible. Although the inter-cluster interference degrades the performance compared to the network-wide CF system, there is a substantial saving because the dimension reduction in each cluster results in significant signaling load and computational cost reduction. In addition, since the number of all users and the scheduled users in each cluster are substantially reduced compared to the network-wide CF, the scheduling costs are significantly reduced as well.

In order to assess the cost of the scheduling methods in
terms of floating point operations (FLOPs), we notice that if the maximum possible number of users ($n=M_c$) are
scheduled, according to the first stage of C-ESG, to obtain $s_1$, $2M_cK_c$ FLOPs are required in each cluster and $2MK$ FLOPs for the CF. We also need 
{$\frac{M_{c}\left ( M_{c}+1 \right )}{2}-1$} FLOPs for all $R_{MMSE}\left ( U_{l-1}\cup \left \{ k \right \} \right )$ during the while loop in each cluster, and
{$\frac{M\left ( M+1 \right )}{2}-1$} FLOPs for CF. For the sum-rate in a cluster, we need $4M_{c}K_{c}^{2}+2K_{c}^{3}+K_{c}$ FLOPs for calculating (\ref{eq:RCL0}), $28M_{c}K_{c}^{2}+14K_{c}^{3}+K_{c}+4$ FLOPs for (\ref{eq:RCL1}), thus, we need $32M_{c}K_{c}^{2}+16K_{c}^{3}+2K_{c}+4$ FLOPs. For calculating the sum-rate in the CF network, we need $K$ flops to compute (\ref{eq:RCF}) and $8MK^{2}+6K^{3}+K$ FLOPs to calculate (\ref{eq:RCF_1}) and therefore, $8MK^{2}+6K^{3}+2K$ FLOPs for $R_{CF}$. Then, for $k_{ex}$ the calculations are done when calculating $s_1$, and for $k_{new}$, we need 
{$2M_{c}\left ( K_{c}-M_{c} \right )$} or 
{$2M\left ( K-M \right )
$} FLOPs for the cluster or CF, respectively. In the other stages ($\textup{stage} \geq 2$), assuming that in each stage we have only one new user, for $k_{ex}$ we need only $2M_c$ or $2M$ FLOPs in cluster or CF, respectively, and for $k_{new}$, the calculations are done in the first stage and we only select the new user. We also require $32M_{c}K_{c}^{2}+16K_{c}^{3}+2K_{c}+4$ FLOPs for $R_{c}$ or $8MK^{2}+6K^{3}+2K$ FLOPs for $R_{CF}$. Accordingly, 
the number of required FLOPs for {a cluster and for} the CF network are as follows, respectively
\vspace{-3mm}
{\begin{multline} \label{NCL}
 N_{cl}=16K_{c}^{4}+16\left ( M_{c}+1\right )K_{c}^{3} \\
+ 2\left ( -16M_{c}^2+16M_{c}+1 \right )K_{c}^{2}\\
+\left ( 4M_{c}+6 \right )K_{c}-\frac{7}{2}M_{c}^2-\frac{7}{2}M_{c}+3
\end{multline}}
\vspace{-6mm}
{\begin{multline}
N_{CF}=6K^{4}+\left ( 2M+6 \right )K^{3}+
2\left ( -4M^{2}+4M+1 \right )K^{2}\\
+\left ( 4M+2 \right )K-\frac{7}{2}M^2+\frac{1}{2}M-1
\end{multline}}
Note that 
{(\ref{NCL}) should be summed over all clusters}.

For a simple comparison of the number of channel parameters, we consider the CF channel coefficient $g_{m,k}$ as shown in section \ref{AA}, which includes 
{1 parameter for} small scale fading coefficient $h_{m,k}$, and 2 parameters for large scale fading coefficient $\beta _{m,k}$ including shadow fading and path loss as described in section \ref{Numerical}.  Then, the number of parameters for channels of a cluster and for CF channels are, respectively,
\vspace{-2mm}
{\begin{equation} \label{LCL}
    L_{cl}=3M_{c}K_{c}
\end{equation}}
\vspace{-5mm}
\begin{equation}
    L_{CF}=3MK
\end{equation}
{For clustered networks (\ref{LCL}) must be summed over all clusters.}
\vspace{-5.7mm}
\section{Numerical Results} \label{Numerical}
\vspace{-1.0mm}
In this section, the performance of C-ESG, ESG (C-ESG for CF networks) and other existing scheduling approaches are assessed in terms of sum-rates. To this end, we have compared the exhaustive search, the greedy and ESG algorithms and the WSR method proposed in \cite{Resource Allocation}. Since the WSR based user scheduling was designed for user-centric clustering, we have adapted it to our scenario so that we can maximize the WSR for the users in a cluster supported by the corresponding APs. 
{We consider $\textbf{G}$ as perfect CSI and model the channel estimate as $\hat{\textbf{G}}=\gamma \textbf{G}$ and the estimation error as $\tilde{\textbf{G}}={\alpha }\textbf{G}$. Thus, the imperfect CSI is modeled as $\textbf{G}_{I}=\hat{\textbf{G}}+\tilde{\textbf{G}}$, where $\alpha ^{2}+\gamma ^{2}=1$, $\gamma > \alpha$ that are selected as $\gamma=\sqrt{0.95}$, $\alpha=\sqrt{0.05}$.}
A squared area with the side length of 400m is considered for the CF network equipped with $M$ randomly located APs. The area includes 
{$K$} users, which are uniformly distributed for simplicity. We have used network-centric clustering with $C=4$ non-overlapping clusters, 
{where cluster $c$ includes} $M_{c}$ randomly located APs and $K_{c}$ uniformly distributed users with uniform power allocation. For statistical robustness, the results are averaged over 10000 trials with different random seeds.
The large scale coefficient in CF channel coefficient is modeled as $\beta _{m,k}=\textup{PL}_{m,k}.10^{\frac{\sigma _{sh}z_{m,k}}{10}}$ where $10^{\frac{\sigma _{sh}z_{m,k}}{10}}$
is the shadow fading with $\sigma _{sh}=8\textup{dB}$, $z_{m,k}\sim\mathcal{N}\left ( 0,1 \right )$, and $\textup{PL}_{m,k}$ is the path loss modeled as \cite{Mobile}
\vspace{-2.5mm}
\begin{equation}
    \textup{PL}_{m,k}=\left\{\begin{matrix}
-\textup{D}-35\log_{10}\left ( d_{m,k} \right ), \textup{if }  d_{m,k}>d_{1}& \\
 -\textup{D}-10\log_{10}\left ( d_{1}^{1.5} d_{m,k}^2\right ), \textup{if }  d_{0}<d_{m,k}\leq d_{1}& \\
 -\textup{D}-10\log_{10}\left ( d_{1}^{1.5} d_{0}^2 \right ), \textup{if }  d_{m,k}\leq d_{0}&
\end{matrix}\right.
\end{equation}
where $d_{m,k}$ is the distance between the $m$th AP and
$k$th user and $\textup{D}$ is
\vspace{-4mm}
\begin{multline}
\textup{D}=46.3+33.9\log_{10}\left ( f \right )-13.82\log_{10}\left ( h_{AP} \right )\\
-\left [ 1.11\log_{10}\left ( f \right )-0.7 \right ]h_{u}+1.56\log_{10}\left ( f \right )-0.8
\end{multline}
{where $f=1900$MHz is the carrier frequency, $h_{AP}=$15m, $h_{u}=$1.5m are the AP and user antenna heights, respectively, $d_{0}=10$m and $d_{1}=$50m. If $d_{m,k}\leq d_{1}$ there is no shadowing.}

Fig.~\ref{fig:schemes, imperfect CSI}a shows the sum-rate performance versus SNR of the clustered CF (CLCF) and CF for different scheduling schemes when the MMSE precoder and imperfect CSI are considered. For all cases, the sum-rate increases with the SNR, however, that C-ESG outperforms other approaches. {We also notice that for each scheduling algorithm, the advantage in information rates of CF over CLCF increases with the SNR because of the additional interference terms in (\ref{eq:RCL1}) for CLCF.} In Fig.~\ref{fig:schemes, imperfect CSI}b, we have considered a network with a small number of users and scheduled up to half of the users, so that we can compare C-ESG with the optimal exhaustive search. We notice that the performance of C-ESG is closer to that of the exhaustive search especially in the CF case.
\begin{figure}
    \centering
    \includegraphics[width=.85\linewidth]{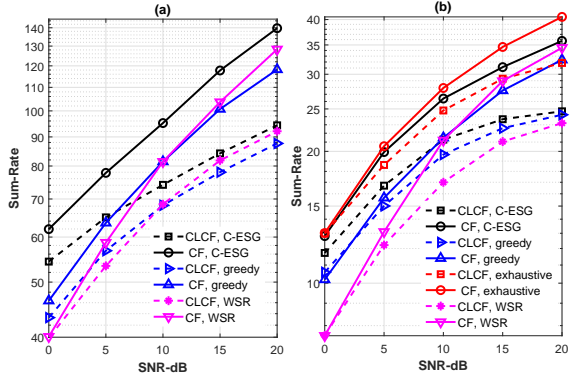}
    \vspace{-0.985em}
    \caption{{{Performance of scheduling schemes for clustered CF (CLCF) and CF, (a): Network with $M=64$, $K=256$ and $n=64$} , (b): Network with $M=64$, $K=16$ and $n=8$ and exhaustive search scheduling included.}}
    \label{fig:schemes, imperfect CSI} \vspace{-2mm}
\end{figure}
With the same ratio between users and APs as considered in Fig.~\ref{fig:schemes, imperfect CSI}b, the number of FLOPs required for user scheduling in networks with different number of APs are shown 
in Fig.~\ref{fig:complexity}a. We can notice that when the size of the network increases, the number of FLOPs also increases. However, the use of clustering resulted in a remarkable decrease in the FLOPs so that for CLCF, it is negligible compared with the CF network. {For a large number of APs, C-ESG has better performance than WSR, which requires less FLOPs}. Fig.~\ref{fig:complexity}b shows that the signaling load in the CLCF is much lower than that of the CF network.
\begin{figure}
    \centering
    \includegraphics[width=.85\linewidth]{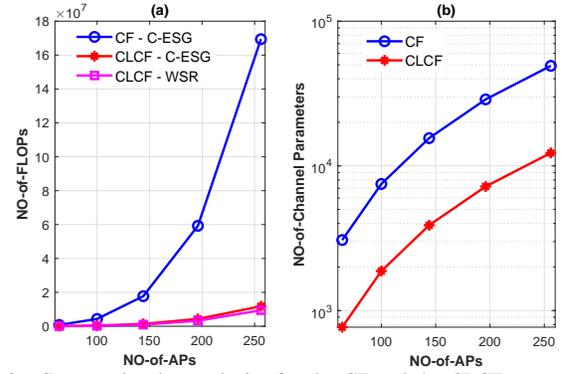}
    \vspace{-0.98em}
    \caption{Computational complexity for the CF and the CLCF networks, (a): Number of required FLOPs for scheduling the users by C-ESG and  WSR methods, (b): Signaling load.}
    \label{fig:complexity}\vspace{-2.5mm}
\end{figure}
{Table~\ref{table:FLOPs1} shows the complexity of the analyzed techniques for CLCF networks. C-ESG requires much less FLOPS than the exhaustive search. Although the complexity of C-ESG is slightly higher, its performance is significantly improved as compared to the greedy and WSR algorithms.}\vspace{-0.5em} 

\begin{table}[htb!]
\begin{small}
\caption{{Computational Complexity of different methods, $M=64$, $K=16$, and $n=8$.}}
\vspace{-1.5em}
\begin{center}
\begin{tabular}{| m{3em} | m{10em} | m{6em} | }
\hline
Network &Scheduling method&NO of FLOPs\\ [0.5ex]
 \hline
\hline
CLCF& C-ESG & 70728 \\
\hline
CLCF& Greedy & 37432 \\
\hline
CLCF& WSR & 52864 \\
\hline
CLCF & Exhaustive search & 221472 \\
\hline
\end{tabular}
\label{table:FLOPs1}
\end{center}
\end{small}
\end{table}

\section{Conclusion}
In this paper, we have proposed the C-ESG multiuser scheduling algorithm, and investigated the performance of network-wide and clustered CF systems in terms of sum-rate, complexity and signaling load. Numerical results illustrate that the proposed C-ESG algorithm shows significant performance improvement in both networks. Moreover, the C-ESG algorithm has a remarkable saving in both the computational cost and signaling load when network clustering is considered.

\end{document}